\documentclass[a4paper,fleqn]{cas-dc}

\usepackage[numbers]{natbib}

\def\tsc#1{\csdef{#1}{\textsc{\lowercase{#1}}\xspace}}
\tsc{WGM}
\tsc{QE}
\tsc{EP}
\tsc{PMS}
\tsc{BEC}
\tsc{DE}

\def\be{\begin{equation}}
	\def\ee{\end{equation}}
\newcommand{\bel}[1]{\begin{eqnarray}\label{#1}}
	\newcommand{\eel}{\end{eqnarray}}
\def\barr{\begin{array}}
	\def\earr{\end{array}}
\def\beq{\begin{eqnarray}}
	\def\eeq{\end{eqnarray}}
\def\bfig{\begin{figure}}
	\def\efig{\end{figure}}

\def\g{\gamma}

\def\c{\chi}
 
\usepackage{amssymb}

\usepackage{graphicx}
\usepackage{amsmath}
\usepackage{subfigure}
\usepackage{bbold}
\usepackage{yfonts}
\usepackage{placeins}
\usepackage{bm} 
\usepackage{nicefrac}
\usepackage{slashed}

\newcommand{\bea}{\begin{eqnarray}}
\newcommand{\eea}{\end{eqnarray}}

\def\S0iU{{\Sigma}^{0i}}

\def\n0{n_{(0)}}
\def\e0{\varepsilon_{(0)}}
\def\P0{P_{(0)}}

\begin{document}
\let\WriteBookmarks\relax
\def\floatpagepagefraction{1}
\def\textpagefraction{.001}
\shorttitle{Equivalence between first-order causal and Israel-Stewart hydrodynamics for systems with constant relaxation times}
\shortauthors{A. Das et~al.}

\title [mode = title]{Equivalence between first-order causal and stable hydrodynamics and Israel-Stewart theory for boost-invariant systems with a constant relaxation time}                      



\author[2]{Arpan Das}[] 

\ead{arpan.das@ifj.edu.pl}

\author[1]{Wojciech Florkowski}[
                       orcid=0000-0002-9215-0238
                        ]
\cormark[1]
\ead{wojciech.florkowski@uj.edu.pl}


\address[1]{Institute of Theoretical Physics, Jagiellonian University,  PL-30-348 Krak\'ow, Poland}

\author[3]{Jorge Noronha}[]

\ead{jn0508@illinois.edu}

\address[3]{Department of Physics,
University of Illinois at Urbana-Champaign, IL 61801-3080, USA}

\author[2]{Radoslaw Ryblewski}[ orcid=0000-0003-3094-7863
                        %
   ]
\ead{radoslaw.ryblewski@ifj.edu.pl}


\address[2]{Institute of Nuclear Physics Polish Academy of Sciences, PL-31-342 Krakow, Poland}



\cortext[cor1]{Corresponding author}


\begin{abstract}
We show that the recently formulated causal and stable first-order hydrodynamics has the same dynamics as Israel-Stewart theory for boost-invariant, Bjorken expanding systems with a conformal equation of state and a regulating sector determined by a constant relaxation time. In this case, the general solution of the new first-order formulation can be determined analytically.  
\end{abstract}



\begin{keywords}
Israel-Stewart theory \sep first-order causal and stable hydrodynamics \sep relativistic heavy-ion collisions
\end{keywords}

\maketitle
\section{Introduction}
Relativistic hydrodynamics has become nowadays the basic theoretical tool for modeling relativistic heavy-ion collisions \cite{Romatschke:2017ejr,Florkowski:2010zz}. It forms the main ingredient of the so-called standard model of such processes, which essentially includes three segments: modeling of the early stage, hydrodynamic description of the space-time evolution of matter, and freeze-out of hadrons \cite{Gale:2013da,Jeon:2015dfa, Jaiswal:2016hex,Busza:2018rrf}. Detailed comparisons of theoretical predictions based on the hydrodynamic approach with the experimental data allow for the determination of various properties of strongly interacting matter such as its equation of state~\cite{Broniowski:2008vp} and kinetic coefficients~\cite{Bozek:2009dw,Noronha-Hostler:2013gga,Bernhard:2019bmu}. The latter include the shear and bulk viscosities. The presence of the shear viscosity affects the response of the hydrodynamic flow to the initial space-time anisotropies of colliding matter~\cite{Romatschke:2007mq,Gale:2012rq}.  

The development of hydrodynamic models for the description of heavy-ion collisions triggered broad studies of formal aspects of hydrodynamics treated as an effective theory describing systems approaching local thermodynamic equilibrium, for a recent review see \cite{Florkowski:2017olj}. Already in the 1970's, it was realized that the relativistic dissipative hydrodynamical formulations derived by Landau and Eckart were not causal~\cite{Israel:1976tn,Hiscock:1983zz,Hiscock:1985zz} and they were replaced by the so-called second order hydrodynamic formalism of Israel and Stewart (IS)~\cite{Israel:1979wp}. The IS theory has been extensively used to describe heavy-ion collisions studied at the Relativistic Heavy Ion Collider (RHIC) at BNL and the Large Hadron Collider (LHC) at CERN. At the same time, more advanced hydrodynamic approaches have been developed, which removed some of the disadvantages of the IS formulation (for example, see \cite{Florkowski:2010cf,Martinez:2010sc,Attems:2018gou,Montenegro:2018bcf,Calzetta:2019dfr}). Formal studies of hydrodynamics have led to very interesting observations such as the asymptotic character of the hydrodynamic gradient expansion~\cite{Heller:2013fn,Denicol:2016bjh,Heller:2016rtz,Grozdanov:2019kge} or the existence of hydrodynamic attractors~\cite{Heller:2015dha,Romatschke:2017vte,Strickland:2017kux,Strickland:2018ayk,Jaiswal:2019cju,Giacalone:2019ldn}. 

IS theory treats the shear stress tensor $\pi^{\mu\nu}$  and the bulk pressure $\Pi$ as independent hydrodynamic variables, in a way similar to the treatment of the local temperature $T(x)$ and the hydrodynamic flow vector $u^\mu(x)$. Only during the space-time evolution of the system $\pi^{\mu\nu}$  and  $\Pi$ may approach their Navier-Stokes values $\pi^{\mu\nu} = 2\eta \sigma^{\mu\nu}$ and $\Pi = -\zeta \partial_\mu u^\mu$ (where $\eta$ and $\zeta$ are the shear and bulk viscosity coefficients, respectively, and $\sigma^{\mu\nu}$ is the shear flow tensor constructed from the derivatives of $u^\mu$). 

Only very recently, a new causal and stable hydrodynamic approach based on a first-order expansion in gradients has been proposed in the works of F.~S.~Bemfica, M.~M.~Disconzi, J.~Noronha, and P.~Kovtun~\cite{Bemfica:2017wps,Bemfica:2019knx,Kovtun:2019hdm}. This approach is based on a more general choice of the hydrodynamic frame and the introduction of a new set of kinetic coefficients that play the role of UV regulators of the theory, which make the theory causal (even in the full nonlinear regime) and linearly stable around equilibrium.   

A natural question that can be asked is how the new formulation (dubbed below shortly as FOCS, for first-order cau-sal and stable) compares to the traditional IS framework. It was shown in Ref.~\cite{Bemfica:2017wps} that the two approaches lead to very similar equations, if applied to boost-invariant conformal systems. In this work we extend this study. We assume that the system's equation of state is conformal but we allow for a non-conformal behavior of the coefficients in the regulating sector of the theory. We show that if the kinetic coefficients are expressed in terms of a constant relaxation time there is an exact match between the dynamics described by FOCS and IS formulations. This allows us to derive the first general analytical solution of the FOCS equations for an expanding system.    

Throughout the paper we use natural units.

\section{Israel-Stewart and first-order stable hydrodynamics} 

The implementation of Israel-Stewart theory undergoing Bjorken flow \cite{Bjorken:1982qr} considered here is reduced to the two equations:
\begin{eqnarray}
\frac{d\varepsilon}{d\tau} &=& -\frac{\varepsilon+p}{\tau} + \frac{\pi}{\tau}, \label{eq:IS1} \\
 \tau_R\frac{d\pi}{d\tau}+\pi &=& \frac{4}{3}\frac{\eta}{\tau} -\left(\frac{4}{3}+\lambda\right)\tau_R\frac{\pi}{\tau},
 \label{eq:IS2}
\end{eqnarray}
where $\varepsilon$ and $p$ are the energy density and pressure, $\pi$ is the rapidity-rapidity component of the shear stress tensor (which should not be mistaken with the bulk pressure, $\Pi$, that is zero in our case), $\eta$ is the shear viscosity coefficient, $\tau_R$ is the relaxation time, and the parameter $\lambda$  \cite{Denicol:2017lxn} is related to the $\tau_{\pi\pi}$ coefficient in the DNMR approach \cite{Denicol:2012cn}. The evolution parameter $\tau = \sqrt{t^2-z^2}$ is the longitudinal proper time. We note that the form of hydrodynamic flow for boost-invariant systems is dictated by symmetry, $u^\mu = (t/\tau,0,0,z/\tau)$, hence it is independent of the choice of the hydrodynamic frame. We also note that all scalar functions depend only on~$\tau$.

For the FOCS approach \cite{Bemfica:2017wps,Kovtun:2019hdm,Bemfica:2019knx}, the evolution equations are reduced to the formula
\begin{eqnarray}
\frac{d\mathcal{E}}{d\tau}+\frac{\mathcal{E}
+\mathcal{P}}{\tau}-\frac{4}{3}\frac{\eta}{\tau^2}=0,
\label{eq:Kovtun}
\end{eqnarray}
where the following constitutive relations are assumed,
\begin{eqnarray}
 \mathcal{E} &=& \epsilon+\varepsilon_1  \frac{dT}{T d\tau}+ \frac{\varepsilon_2}{\tau},\nonumber\\
 \mathcal{P} &=& p +\pi_1 \frac{dT}{T d\tau}+ \frac{\pi_2}{\tau}.
 \label{eq:Kovtuncoef}
 \end{eqnarray}
Here we have used the properties $ \partial_\mu u^\mu = 1/\tau$ and $u^\mu \partial_\mu f(\tau) = df/d\tau$, where $f$ is an arbitrary function of the proper time $\tau$.

Assuming the conformal equation of state
\begin{eqnarray}
p=\frac{1}{3}\varepsilon=\frac{aT^4}{3},
\end{eqnarray}
where $a$ is a constant (usually proportional to the number of internal degrees of freedom of the particles forming a fluid) and $T$ is the temperature, we rewrite Eq.~(\ref{eq:IS1}) as 
\begin{eqnarray}
\frac{dT}{d\tau}=\frac{\pi}{4aT^3\tau}-\frac{T}{3\tau}.
\label{eq:IS1T}
\end{eqnarray}
Introducing the variable
\begin{eqnarray}
 y= \frac{dT}{d\tau}
 \label{eq:y}
\end{eqnarray} 
and taking the derivative of Eq.~(\ref{eq:IS1T}) with respect to $\tau$,  we obtain
\begin{eqnarray}
  \frac{d\pi}{d\tau} = 4aT^3y+4aT^3\tau \frac{dy}{d\tau}+12 aT^2\tau y^2+\frac{16}{3}a T^3 y.
 \end{eqnarray}
This allows us to rewrite Eq.~(\ref{eq:IS2}) as
\begin{eqnarray}
&&   4a \tau_R T^3\frac{dy}{d\tau} + 12\tau_R aT^2 y^2 \nonumber \\
&& + aT^3 y\bigg[4 +\bigg(\frac{28}{3}+4\bigg(\frac{4}{3}+\lambda\bigg)\bigg)\frac{\tau_R}{\tau}\bigg] \nonumber \\
&& +\frac{4 aT^4}{3 \tau} + \frac{4}{3}aT^4 \bigg(\frac{4}{3}
+\lambda\bigg)\frac{\tau_R}{\tau^2}-\frac{4}{3}\frac{\eta}{\tau^2}=0. 
\label{eq:dydt}
 \end{eqnarray}
 Equations (\ref{eq:y}) and (\ref{eq:dydt}) are coupled differential equations for the functions $T$ and $y$, which are completely equivalent to the original IS equations. We note that Eq.~(\ref{eq:dydt}) has the form of a Ricatti equation ($a y'+ b y^2 + c y + d = 0$, with $b/a \ne 0$ and $c/a \ne 0$), which was analyzed recently in more detail in \cite{Denicol:2017lxn}.

\section{Regulating sector in FOCS} 

Let us discuss in more detail the regulating sector of the FOCS approach. In natural units, the coefficient functions $\varepsilon_1$, $\varepsilon_2$, $\pi_1$, and $\pi_2$ have dimension of energy cubed, so for conformal systems they should scale as $T^3$. Similarly, in this case the IS relaxation time $\tau_R$ should be inversely proportional to $T$, while $\eta$ should scale with $T^3$, yielding a dimensionless ratio of the shear viscosity $\eta$ to the entropy density $s$. For strictly conformal systems one requires also that  $\mathcal{E} = 3 \mathcal{P}$, which implies that the $\pi_i$ coefficients are one third of the $\varepsilon_i$ coefficients. This choice has been made in \cite{Bemfica:2017wps}, with an additional constraint that $\varepsilon_1 = 3 \varepsilon_2$ to ensure invariance under Weyl transformations \cite{Bemfica:2017wps}. 

In this work we want to discuss yet another case, where the coefficients $\varepsilon_1$, $\varepsilon_2$, $\pi_1$, and $\pi_2$ are expressed in terms of a constant relaxation time. This leads to parametrizations of the type $x_i = x_i^0 T^4$, where $x_i$ stands for any of the FOCS coefficients mentioned above and $x^0_i$ has dimension of time (fm). We think that this assumption is interesting from the point of view where the terms containing  $\varepsilon_1$, $\varepsilon_2$, $\pi_1$, and $\pi_2$ are interpreted as ultraviolet regulators. In general, the regularization or renormalization procedure of a classically scale-invariant theory introduces an energy scale, as it happens in the case of pure Yang-Mills theory (known exceptions of this rule include, for instance, $\mathcal{N}=4$ supersymmetric Yang-Mills theory). 

To include and discuss different cases together we rewrite Eq.~(\ref{eq:Kovtuncoef}) as
\begin{eqnarray}
 \mathcal{E} &=& a T^4 + \varepsilon^0_1 \,T^{n} \, \frac{dT}{T d\tau}+ \frac{\varepsilon^0_2}{\tau} \,T^n,
 \nonumber\\
 \mathcal{P} &=& \frac{aT^4}{3} +\pi^0_1 \,T^{n} \, \frac{dT}{T d\tau}+ 
 \frac{\pi^0_2}{\tau} \,T^n,
 \label{eq:Kovtuncoef1}
 \end{eqnarray}
where $\varepsilon^0_1$, $\varepsilon^0_2$, $\pi^0_1$, and $\pi^0_2$ are dimensionless ($n=3$) or dimensionful ($n \neq 3$) constants. The power $n$ can take different values depending on the case we want to discuss. Substituting (\ref{eq:Kovtuncoef1}) into Eq.~(\ref{eq:Kovtun}), and using Eq.~(\ref{eq:y}), we find
\begin{eqnarray}
&& \varepsilon^0_1 T^{n-1}\frac{dy}{d\tau} + (n-1) \, \varepsilon^0_1 \,T^{n-2}\,
y^2
\nonumber \\
&&
+\bigg(4aT^3 + (\varepsilon^0_1+\pi^0_1 + n\, \varepsilon^0_2)
\frac{T^{n-1} }{\tau} \, 
\bigg)y \nonumber \\
&& +\frac{4}{3\tau}aT^4 +\frac{\pi_2\,T^n}{\tau^2}
-\frac{4}{3}\frac{\eta}{\tau^2}=0.
\label{eq:Kovt2nd1}
\end{eqnarray}
Equations (\ref{eq:y}) and (\ref{eq:Kovt2nd1}) are coupled first-order differential equations that can be treated as the basis of the FOCS formulation in our setup.

\section{Comparison between the two frameworks} 

Using the parametrizations defined above we can formulate the IS and FOCS frameworks in terms of the two differential equations for the temperature $T$ and its derivative $y=dT/d\tau$. A natural question is if these two formulations, when written in this form, are actually identical describing thus the same dynamics. Since Eq.~(\ref{eq:y}) is common for the two approaches, one simply has to check if Eqs.~(\ref{eq:dydt}) and (\ref{eq:Kovt2nd1}) are equivalent. After equating the terms with the same derivatives of the function $y$ in Eq.~(\ref{eq:dydt}) and (\ref{eq:Kovt2nd1}) we find:
\begin{eqnarray}
 \varepsilon^0_1 &=& 4a\tau_R T^{4-n},\label{eq:w2} \\
 \varepsilon^0_1 &=&\frac{12}{n-1} a\tau_R T^{4-n},\label{eq:w12}\\
 \pi^0_1 &=& \frac{4}{3} a \tau_R (11 + 3 \lambda) T^{4-n} - \varepsilon^0_1 - n \varepsilon_2^0,\label{eq:w11} \\
 \pi^0_2 &=& \frac{4}{9} a \tau_R
 \bigg(4 +3 \lambda\bigg) T^{4-n}.
 \label{eq:w0}
\end{eqnarray}

One can easily notice that in the strictly conformal case,  $n=3$, it is impossible to exactly match the FOCS and IS equations, even though the evolution equation for $y$ in both formulations can be written as a Ricatti equation. The param-etrization of the Ricatti equation found in the conformal case in \cite{Bemfica:2017wps} uses a function $\chi$ that is related to our parametrization through the formula
\begin{eqnarray}
\chi = \frac{1}{3} \varepsilon_1^0 T^3 = \varepsilon_2^0 T^3.
\end{eqnarray}
Moreover, in \cite{Bemfica:2017wps} one uses the relation $\pi_1^0 = (1/3) \varepsilon_1^0 = \varepsilon_2^0$.

A very interesting situation takes place when $n=4$. In this case Eqs.~(\ref{eq:w2}) and (\ref{eq:w12}) are fully consistent and the kinetic coefficient $\varepsilon_1^0$ has dimension of fm and, thus, it can be treated as a fixed relaxation time related to $\tau_R$ (which is also constant). Equations~(\ref{eq:w11}) and (\ref{eq:w0}) determine the values of $\pi_1^0$ and $\pi_2^0$ in terms of the IS relaxation time,  $\varepsilon_1^0$ and $\varepsilon_2^0$. Although for $n=4$ the system of equations (\ref{eq:w2})--(\ref{eq:w0}) can be adjusted to exactly match the IS equations, at first it seems that the matching is underdetermined as only the sum $\pi_1^0 + 4 \varepsilon_2^0$ is constrained by Eq.~(\ref{eq:w11}). We discuss this apparent issue in more detail below. 

\subsection{Bulk viscosity constraint}
\label{sec:bulk}

Further insights about the identification of the FOCS and IS approaches can be gained from the analysis of the bulk viscosity coefficient $\zeta$, which should vanish for systems with a conformal equation of state. In the FOCS approach, the bulk viscosity appears as a linear combination of the regulators and one can show that \cite{Kovtun:2019hdm,Bemfica:2019knx}
\begin{eqnarray}
\zeta = \chi_3 - \chi_4 + c_s^2 (\chi_2-\chi_1).
\label{eq:zeta1}
\end{eqnarray}
Above, $c_s^2 = 1/3$ is the speed of sound squared. The coefficients $\chi_i$ appearing in (\ref{eq:zeta1}) can be directly related to our parametrizations through the following relations ($n=4$):
\begin{eqnarray}
3 \chi_1 = \varepsilon_1^0 T^4,~  
\chi_2 = \varepsilon_2^0 T^4,~
3 \chi_3 = \pi_1^0 T^4,~
\chi_4 = \pi_2^0 T^4.
\label{eq:chiepspi}
\end{eqnarray}
Hence, the requirement that the bulk viscosity vanishes implies that 
\begin{eqnarray}
\pi_1^0 - 3 \pi_2^0 + \varepsilon_2^0 - \frac{\varepsilon_1^0}{3}  = 0.
\label{eq:zeta0}
\end{eqnarray}
The constraint (\ref{eq:zeta0}) leads to the following expressions: 
\begin{eqnarray}
 \varepsilon^0_1 &=& 4a\tau_R,\label{eq:1} \\
 \varepsilon^0_2 &=&\frac{4}{3} a\tau_R,\label{eq:2}\\
 \pi^0_1 &=& \frac{4}{3} a \tau_R (4 + 3\lambda),\label{eq:3} \\
 \pi^0_2 &=& \frac{4}{9} a \tau_R
 \bigg(4 +3 \lambda\bigg).
 \label{eq:4}
\end{eqnarray}
In terms of the coefficients used in \cite{Bemfica:2019knx}, the relations above imply that $\chi_1=\chi_2$ and $\chi_3 = \chi_4$. Therefore, one can see that the condition of vanishing bulk viscosity removes the apparent ambiguity in the mapping present in Eq.~(\ref{eq:w11}).

\subsection{Constraints from causality and linear stability in the FOCS formulation}

For the type of Israel-Stewart theory considered here, causality and stability around equilibrium hold when $\eta/(s\tau_R T)$ $\leq 1/2$ (where $s=4\varepsilon/3T$) \cite{Pu:2009fj}. We note that this is a statement obtained after linearizing the equations  around equilibrium and, thus, no constraint is known for the $\lambda$ coefficient, as it does not contribute in a linearized analysis. However, this coefficient is known in the 14-moment approximation to be equal to 10/21 \cite{Denicol:2012cn}, while the shear viscosity is given by $\eta = 4 \varepsilon \tau_R/15$.   

Ref.\ \cite{Bemfica:2019knx} derived conditions for the transport coefficients in the FOCS approach that ensure causality in the full nonlinear regime. Also, linear stability conditions around equilibrium were discussed in both \cite{Bemfica:2019knx} and \cite{Kovtun:2019hdm}. We refer the reader to Section III A and B of Ref.\ \cite{Bemfica:2019knx} for the set of inequalities that must be fulfilled for causality and stability to hold in the FOCS approach. 

Since the causality conditions in the full nonlinear regime are known for the first-order approach, it is interesting to consider if the identifications made in \eqref{eq:1}-\eqref{eq:4} can fulfill these conditions when the 14-moment values for $\eta$, $\tau_R$, and $\lambda$ are used. In this case, from \eqref{eq:1}-\eqref{eq:4} one finds that $\chi_1 =\chi_2=5 \eta$ and $\chi_3=\chi_4 = 5\eta (4+3\lambda)/3$, which can then be plugged in the conditions stated in \cite{Bemfica:2019knx}. One can show that causality in the FOCS theory is violated if one uses the 14-moment value $\lambda=10/21$. Causality in the FOCS approach with coefficients given by \eqref{eq:1}-\eqref{eq:4} can be fulfilled, however, if $\lambda$ is negative. For instance, all the conditions for causality and stability are satisfied if $\lambda = -2\eta/\lambda_{BDN}-7/15$, where $\lambda_{BDN}\geq \eta$ stands for the heat flow-like transport coefficient present in the FOCS approach (using the conventions in \cite{Bemfica:2019knx}).

 In any case, it is not currently known if the 14-moment value for $\lambda$ in IS theory leads to causality violations once the full nonlinear dynamics of the equations is taken into account. The results presented herein may suggest that the IS parameter $\lambda=10/21$ can be at odds with causality when one goes beyond the linearized regime. However, such a conjecture can only be checked once a full nonlinear analysis of causality in Israel-Stewart theory, performed under general conditions, is available. So far, such general statements about causality in the nonlinear regime of Israel-Stewart theory have  been obtained in \cite{Bemfica:2019cop} in the case where only bulk viscosity (i.e., no shear or particle diffusion effects) is taken into account (a nonlinear study involving shear and bulk viscosities in IS theory under strong symmetry conditions can be found in \cite{Floerchinger:2017cii}).

\subsection{General analytical solution for the FOCS approach matched to IS theory}

Ref.\ \cite{Denicol:2017lxn} found the general solution of the IS equations in \eqref{eq:IS1} and \eqref{eq:IS2} in Bjorken flow. The analytical expressions for $\varepsilon(\tau)$ and $\pi(\tau)$ can be obtained from Eqs.\ (15) and (16) of \cite{Denicol:2017lxn}. The matching to IS theory worked out in this paper (see \eqref{eq:1}-\eqref{eq:4}) implies that the general solution for the energy density in IS found in Eq.\ (15) of \cite{Denicol:2017lxn} also holds for the FOCS theory (for appropriate values of $\lambda$). Therefore, under these conditions, the general solution for the energy density in the FOCS approach with a constant relaxation time as defined here is
\begin{align}
  & \varepsilon(\hat\tau) = \varepsilon _{0}\left( \frac{\hat{\tau}_{0}}{\hat{\tau}}\right) ^{\frac{4%
}{3}+\frac{\lambda +1}{2}}\,\exp \left( -\frac{\hat{\tau}-\hat{\tau}_{0}}{2}%
\right) \nonumber \\ 
&~~\times\left[ \frac{M_{-\frac{\lambda +1}{2},\frac{\sqrt{\lambda ^{2}+4a}}{2%
}}(\hat{\tau})+\alpha \,W_{-\frac{\lambda +1}{2},\frac{\sqrt{\lambda ^{2}+4a}%
}{2}}(\hat{\tau})}{M_{-\frac{\lambda +1}{2},\frac{\sqrt{\lambda ^{2}+4a}}{2}%
}(\hat{\tau}_{0})+\alpha \,W_{-\frac{\lambda +1}{2},\frac{\sqrt{\lambda
^{2}+4a}}{2}}(\hat{\tau}_{0})}\right] 
    \end{align}
    where $\hat\tau = \tau/\tau_R$, $\hat\tau_0$ is the initial time, $\varepsilon_0$ and $\alpha$ are constants that define the initial value problem, and $M_{k,\mu}(x)$ and $W_{k,\mu}(x)$ are Whittaker functions. This is the first analytical solution of the viscous relativistic hydrodynamics equations derived from the new first-order approach put forward in Refs.\ \cite{Bemfica:2017wps,Bemfica:2019knx,Kovtun:2019hdm}. It should be clear also that the mapping between these approaches found here immediately establishes the properties of the hydrodynamic attractor in the FOCS approach in this case, as they can be extracted from the analysis already performed in IS theory in \cite{Denicol:2017lxn}.

\section{Conclusions}

In this work we have compared the recent first-order causal and stable formulation of relativistic hydrodynamics with conventional Israel-Stewart theory. To make such a comparison feasible, we have restricted our study to boost-invariant, baryon-free  systems with a conformal equation of state. In the strictly conformal case, where the regulator sectors of the theories are also determined from conformal invariance, the two approaches cannot be exactly matched, although they are based on the same system of differential equations (see Ref.\ \cite{Bemfica:2017wps}). If the regulator sectors of the theories are determined by a constant relaxation time, there exists a mapping between the FOCS and IS approaches that makes their dynamics exactly the same. This implies that one can use the results in \cite{Denicol:2017lxn} to determine the first general analytical solution of the FOCS equations of motion, as we showed in this paper. The causality conditions for the FOCS approach found in \cite{Bemfica:2019knx} proved to be relevant when determining the range of acceptable values of the transport coefficients in the FOCS approach, after the matching to IS theory. In fact, we showed that this matching to IS theory is only well defined if the IS parameter $\lambda$ takes values that are distinct from the standard 14-moment result.       

Our results help to clarify mutual relations between FOCS and more traditional formulations of relativistic dissipative hydrodynamics. Further investigations of more general systems are of course mandatory in this respect. Although for more complex system simple relations connecting FOCS with second order hydrodynamic frameworks may not exist (since FOCS yields four second-order equations which are in general equivalent to eight first-order equations, while IS is based on ten equations describing the time evolution of ten independent components of the symmetric energy-momentum tensor), it is in our opinion very interesting to identify the cases where such constructions are possible. This helps to better understand the physics behind this new first-order formulation, which may eventually become an attractive alternative to more traditional hydrodynamic frameworks.

\medskip
{\it Acknowledgements.} WF, RR, and JN would like to thank the participants of the Banff Workshop on Theoretical Foundations of Relativistic Hydrodynamics (19w5048) for inspiring discussions during the meeting, which triggered these investigations. The work of WF, RR, and AD was supported in part by the Polish National Science Center Grants No.~2016\\/23/B/ST2/00717 and No.~2018/30/E/ST2/00432.

\printcredits

\bibliographystyle{utphys} 


\begin{thebibliography}{41}
\providecommand{\natexlab}[1]{#1}
\providecommand{\url}[1]{\texttt{#1}}
\providecommand{\href}[2]{#2}
\providecommand{\path}[1]{#1}
\providecommand{\DOIprefix}{doi:}
\providecommand{\ArXivprefix}{arXiv:}
\providecommand{\URLprefix}{URL: }
\providecommand{\Pubmedprefix}{pmid:}
\providecommand{\doi}[1]{\href{http://dx.doi.org/#1}{\path{#1}}}
\providecommand{\Pubmed}[1]{\href{pmid:#1}{\path{#1}}}
\providecommand{\BIBand}{and}
\providecommand{\bibinfo}[2]{#2}
\ifx\xfnm\undefined \def\xfnm[#1]{\unskip,\space#1}\fi
\makeatletter\def\@biblabel#1{#1.}\makeatother
\bibitem[{Romatschke and Romatschke(2019)}]{Romatschke:2017ejr}
\bibinfo{author}{Romatschke\xfnm[ P.]}, \bibinfo{author}{Romatschke\xfnm[ U.]}.
\newblock \bibinfo{title}{{Relativistic Fluid Dynamics In and Out of
  Equilibrium}}.
\newblock Cambridge Monographs on Mathematical Physics;
  \bibinfo{publisher}{Cambridge University Press}; \bibinfo{year}{2019}.
\newblock ISBN \bibinfo{isbn}{9781108483681, 9781108750028}.
\newblock \DOIprefix\doi{10.1017/9781108651998}.
  \href{http://arxiv.org/abs/1712.05815}{\tt arXiv:1712.05815}.
\bibitem[{Florkowski(2010)}]{Florkowski:2010zz}
\bibinfo{author}{Florkowski\xfnm[ W.]}.
\newblock \bibinfo{title}{{Phenomenology of Ultra-Relativistic Heavy-Ion
  Collisions}}.
\newblock \bibinfo{year}{2010}.
\newblock ISBN \bibinfo{isbn}{9789814280662}.
\bibitem[{Gale et~al.(2013{\natexlab{a}})Gale, Jeon and Schenke}]{Gale:2013da}
\bibinfo{author}{Gale\xfnm[ C.]}, \bibinfo{author}{Jeon\xfnm[ S.]},
  \bibinfo{author}{Schenke\xfnm[ B.]}.
\newblock \bibinfo{title}{{Hydrodynamic Modeling of Heavy-Ion Collisions}}.
\newblock \emph{\bibinfo{journal}{Int J Mod Phys}}
  \bibinfo{year}{2013}{\natexlab{a}};\bibinfo{volume}{A28}:\bibinfo{pages}{1340011}.
\newblock \DOIprefix\doi{10.1142/S0217751X13400113}.
  \href{http://arxiv.org/abs/1301.5893}{\tt arXiv:1301.5893}.
\bibitem[{Jeon and Heinz(2015)}]{Jeon:2015dfa}
\bibinfo{author}{Jeon\xfnm[ S.]}, \bibinfo{author}{Heinz\xfnm[ U.]}.
\newblock \bibinfo{title}{{Introduction to Hydrodynamics}}.
\newblock \emph{\bibinfo{journal}{Int J Mod Phys}}
  \bibinfo{year}{2015};\bibinfo{volume}{E24}(\bibinfo{number}{10}):\bibinfo{pages}{1530010}.
\newblock \DOIprefix\doi{10.1142/S0218301315300106}.
  \href{http://arxiv.org/abs/1503.03931}{\tt arXiv:1503.03931}.
\bibitem[{Jaiswal and Roy(2016)}]{Jaiswal:2016hex}
\bibinfo{author}{Jaiswal\xfnm[ A.]}, \bibinfo{author}{Roy\xfnm[ V.]}.
\newblock \bibinfo{title}{{Relativistic hydrodynamics in heavy-ion collisions:
  general aspects and recent developments}}.
\newblock \emph{\bibinfo{journal}{Adv High Energy Phys}}
  \bibinfo{year}{2016};\bibinfo{volume}{2016}:\bibinfo{pages}{9623034}.
\newblock \DOIprefix\doi{10.1155/2016/9623034}.
  \href{http://arxiv.org/abs/1605.08694}{\tt arXiv:1605.08694}.
\bibitem[{Busza et~al.(2018)Busza, Rajagopal and van~der Schee}]{Busza:2018rrf}
\bibinfo{author}{Busza\xfnm[ W.]}, \bibinfo{author}{Rajagopal\xfnm[ K.]},
  \bibinfo{author}{van~der Schee\xfnm[ W.]}.
\newblock \bibinfo{title}{{Heavy Ion Collisions: The Big Picture, and the Big
  Questions}}.
\newblock \emph{\bibinfo{journal}{Ann Rev Nucl Part Sci}}
  \bibinfo{year}{2018};\bibinfo{volume}{68}:\bibinfo{pages}{339--376}.
\newblock \DOIprefix\doi{10.1146/annurev-nucl-101917-020852}.
  \href{http://arxiv.org/abs/1802.04801}{\tt arXiv:1802.04801}.
\bibitem[{Broniowski et~al.(2008)Broniowski, Chojnacki, Florkowski and
  Kisiel}]{Broniowski:2008vp}
\bibinfo{author}{Broniowski\xfnm[ W.]}, \bibinfo{author}{Chojnacki\xfnm[ M.]},
  \bibinfo{author}{Florkowski\xfnm[ W.]}, \bibinfo{author}{Kisiel\xfnm[ A.]}.
\newblock \bibinfo{title}{{Uniform Description of Soft Observables in Heavy-Ion
  Collisions at s(NN)**(1/2) = 200 GeV**2}}.
\newblock \emph{\bibinfo{journal}{Phys Rev Lett}}
  \bibinfo{year}{2008};\bibinfo{volume}{101}:\bibinfo{pages}{022301}.
\newblock \DOIprefix\doi{10.1103/PhysRevLett.101.022301}.
  \href{http://arxiv.org/abs/0801.4361}{\tt arXiv:0801.4361}.
\bibitem[{Bozek(2010)}]{Bozek:2009dw}
\bibinfo{author}{Bozek\xfnm[ P.]}.
\newblock \bibinfo{title}{{Bulk and shear viscosities of matter created in
  relativistic heavy-ion collisions}}.
\newblock \emph{\bibinfo{journal}{Phys Rev}}
  \bibinfo{year}{2010};\bibinfo{volume}{C81}:\bibinfo{pages}{034909}.
\newblock \DOIprefix\doi{10.1103/PhysRevC.81.034909}.
  \href{http://arxiv.org/abs/0911.2397}{\tt arXiv:0911.2397}.
\bibitem[{Noronha-Hostler et~al.(2013)Noronha-Hostler, Denicol, Noronha,
  Andrade and Grassi}]{Noronha-Hostler:2013gga}
\bibinfo{author}{Noronha-Hostler\xfnm[ J.]}, \bibinfo{author}{Denicol\xfnm[
  G.S.]}, \bibinfo{author}{Noronha\xfnm[ J.]}, \bibinfo{author}{Andrade\xfnm[
  R.P.G.]}, \bibinfo{author}{Grassi\xfnm[ F.]}.
\newblock \bibinfo{title}{{Bulk Viscosity Effects in Event-by-Event
  Relativistic Hydrodynamics}}.
\newblock \emph{\bibinfo{journal}{Phys Rev}}
  \bibinfo{year}{2013};\bibinfo{volume}{C88}(\bibinfo{number}{4}):\bibinfo{pages}{044916}.
\newblock \DOIprefix\doi{10.1103/PhysRevC.88.044916}.
  \href{http://arxiv.org/abs/1305.1981}{\tt arXiv:1305.1981}.
\bibitem[{Bernhard et~al.(2019)Bernhard, Moreland and Bass}]{Bernhard:2019bmu}
\bibinfo{author}{Bernhard\xfnm[ J.E.]}, \bibinfo{author}{Moreland\xfnm[ J.S.]},
  \bibinfo{author}{Bass\xfnm[ S.A.]}.
\newblock \bibinfo{title}{{Bayesian estimation of the specific shear and bulk
  viscosity of quark–gluon plasma}}.
\newblock \emph{\bibinfo{journal}{Nature Phys}}
  \bibinfo{year}{2019};\bibinfo{volume}{15}(\bibinfo{number}{11}):\bibinfo{pages}{1113--1117}.
\newblock \DOIprefix\doi{10.1038/s41567-019-0611-8}.
\bibitem[{Romatschke and Romatschke(2007)}]{Romatschke:2007mq}
\bibinfo{author}{Romatschke\xfnm[ P.]}, \bibinfo{author}{Romatschke\xfnm[ U.]}.
\newblock \bibinfo{title}{{Viscosity Information from Relativistic Nuclear
  Collisions: How Perfect is the Fluid Observed at RHIC?}}
\newblock \emph{\bibinfo{journal}{Phys Rev Lett}}
  \bibinfo{year}{2007};\bibinfo{volume}{99}:\bibinfo{pages}{172301}.
\newblock \DOIprefix\doi{10.1103/PhysRevLett.99.172301}.
  \href{http://arxiv.org/abs/0706.1522}{\tt arXiv:0706.1522}.
\bibitem[{Gale et~al.(2013{\natexlab{b}})Gale, Jeon, Schenke, Tribedy and
  Venugopalan}]{Gale:2012rq}
\bibinfo{author}{Gale\xfnm[ C.]}, \bibinfo{author}{Jeon\xfnm[ S.]},
  \bibinfo{author}{Schenke\xfnm[ B.]}, \bibinfo{author}{Tribedy\xfnm[ P.]},
  \bibinfo{author}{Venugopalan\xfnm[ R.]}.
\newblock \bibinfo{title}{{Event-by-event anisotropic flow in heavy-ion
  collisions from combined Yang-Mills and viscous fluid dynamics}}.
\newblock \emph{\bibinfo{journal}{Phys Rev Lett}}
  \bibinfo{year}{2013}{\natexlab{b}};\bibinfo{volume}{110}(\bibinfo{number}{1}):\bibinfo{pages}{012302}.
\newblock \DOIprefix\doi{10.1103/PhysRevLett.110.012302}.
  \href{http://arxiv.org/abs/1209.6330}{\tt arXiv:1209.6330}.
\bibitem[{Florkowski et~al.(2018)Florkowski, Heller and
  Spalinski}]{Florkowski:2017olj}
\bibinfo{author}{Florkowski\xfnm[ W.]}, \bibinfo{author}{Heller\xfnm[ M.P.]},
  \bibinfo{author}{Spalinski\xfnm[ M.]}.
\newblock \bibinfo{title}{{New theories of relativistic hydrodynamics in the
  LHC era}}.
\newblock \emph{\bibinfo{journal}{Rept Prog Phys}}
  \bibinfo{year}{2018};\bibinfo{volume}{81}(\bibinfo{number}{4}):\bibinfo{pages}{046001}.
\newblock \DOIprefix\doi{10.1088/1361-6633/aaa091}.
  \href{http://arxiv.org/abs/1707.02282}{\tt arXiv:1707.02282}.
\bibitem[{Israel(1976)}]{Israel:1976tn}
\bibinfo{author}{Israel\xfnm[ W.]}.
\newblock \bibinfo{title}{{Nonstationary irreversible thermodynamics: A Causal
  relativistic theory}}.
\newblock \emph{\bibinfo{journal}{Annals Phys}}
  \bibinfo{year}{1976};\bibinfo{volume}{100}:\bibinfo{pages}{310--331}.
\newblock \DOIprefix\doi{10.1016/0003-4916(76)90064-6}.
\bibitem[{Hiscock and Lindblom(1983)}]{Hiscock:1983zz}
\bibinfo{author}{Hiscock\xfnm[ W.A.]}, \bibinfo{author}{Lindblom\xfnm[ L.]}.
\newblock \bibinfo{title}{{Stability and causality in dissipative relativistic
  fluids}}.
\newblock \emph{\bibinfo{journal}{Annals Phys}}
  \bibinfo{year}{1983};\bibinfo{volume}{151}:\bibinfo{pages}{466--496}.
\newblock \DOIprefix\doi{10.1016/0003-4916(83)90288-9}.
\bibitem[{Hiscock and Lindblom(1985)}]{Hiscock:1985zz}
\bibinfo{author}{Hiscock\xfnm[ W.A.]}, \bibinfo{author}{Lindblom\xfnm[ L.]}.
\newblock \bibinfo{title}{{Generic instabilities in first-order dissipative
  relativistic fluid theories}}.
\newblock \emph{\bibinfo{journal}{Phys Rev}}
  \bibinfo{year}{1985};\bibinfo{volume}{D31}:\bibinfo{pages}{725--733}.
\newblock \DOIprefix\doi{10.1103/PhysRevD.31.725}.
\bibitem[{Israel and Stewart(1979)}]{Israel:1979wp}
\bibinfo{author}{Israel\xfnm[ W.]}, \bibinfo{author}{Stewart\xfnm[ J.M.]}.
\newblock \bibinfo{title}{{Transient relativistic thermodynamics and kinetic
  theory}}.
\newblock \emph{\bibinfo{journal}{Annals Phys}}
  \bibinfo{year}{1979};\bibinfo{volume}{118}:\bibinfo{pages}{341--372}.
\newblock \DOIprefix\doi{10.1016/0003-4916(79)90130-1}.
\bibitem[{Florkowski and Ryblewski(2011)}]{Florkowski:2010cf}
\bibinfo{author}{Florkowski\xfnm[ W.]}, \bibinfo{author}{Ryblewski\xfnm[ R.]}.
\newblock \bibinfo{title}{{Highly-anisotropic and strongly-dissipative
  hydrodynamics for early stages of relativistic heavy-ion collisions}}.
\newblock \emph{\bibinfo{journal}{Phys Rev}}
  \bibinfo{year}{2011};\bibinfo{volume}{C83}:\bibinfo{pages}{034907}.
\newblock \DOIprefix\doi{10.1103/PhysRevC.83.034907}.
  \href{http://arxiv.org/abs/1007.0130}{\tt arXiv:1007.0130}.
\bibitem[{Martinez and Strickland(2010)}]{Martinez:2010sc}
\bibinfo{author}{Martinez\xfnm[ M.]}, \bibinfo{author}{Strickland\xfnm[ M.]}.
\newblock \bibinfo{title}{{Dissipative Dynamics of Highly Anisotropic
  Systems}}.
\newblock \emph{\bibinfo{journal}{Nucl Phys}}
  \bibinfo{year}{2010};\bibinfo{volume}{A848}:\bibinfo{pages}{183--197}.
\newblock \DOIprefix\doi{10.1016/j.nuclphysa.2010.08.011}.
  \href{http://arxiv.org/abs/1007.0889}{\tt arXiv:1007.0889}.
\bibitem[{Attems et~al.(2018)Attems, Bea, Casalderrey-Solana, Mateos, Triana
  and Zilhão}]{Attems:2018gou}
\bibinfo{author}{Attems\xfnm[ M.]}, \bibinfo{author}{Bea\xfnm[ Y.]},
  \bibinfo{author}{Casalderrey-Solana\xfnm[ J.]}, \bibinfo{author}{Mateos\xfnm[
  D.]}, \bibinfo{author}{Triana\xfnm[ M.]}, \bibinfo{author}{Zilhão\xfnm[
  M.]}.
\newblock \bibinfo{title}{{Holographic Collisions across a Phase Transition}}.
\newblock \emph{\bibinfo{journal}{Phys Rev Lett}}
  \bibinfo{year}{2018};\bibinfo{volume}{121}(\bibinfo{number}{26}):\bibinfo{pages}{261601}.
\newblock \DOIprefix\doi{10.1103/PhysRevLett.121.261601}.
  \href{http://arxiv.org/abs/1807.05175}{\tt arXiv:1807.05175}.
\bibitem[{Montenegro and Torrieri(2018)}]{Montenegro:2018bcf}
\bibinfo{author}{Montenegro\xfnm[ D.]}, \bibinfo{author}{Torrieri\xfnm[ G.]}.
\newblock \bibinfo{title}{{Causality and dissipation in relativistic
  polarizeable fluids}}
  \bibinfo{year}{2018};\href{http://arxiv.org/abs/1807.02796}{\tt
  arXiv:1807.02796}.
\bibitem[{Calzetta and Cantarutti(2019)}]{Calzetta:2019dfr}
\bibinfo{author}{Calzetta\xfnm[ E.]}, \bibinfo{author}{Cantarutti\xfnm[ L.]}.
\newblock \bibinfo{title}{{Dissipative type theories for Bjorken and Gubser
  flows}} \bibinfo{year}{2019};\href{http://arxiv.org/abs/1912.10562}{\tt
  arXiv:1912.10562}.
\bibitem[{Heller et~al.(2013)Heller, Janik and Witaszczyk}]{Heller:2013fn}
\bibinfo{author}{Heller\xfnm[ M.P.]}, \bibinfo{author}{Janik\xfnm[ R.A.]},
  \bibinfo{author}{Witaszczyk\xfnm[ P.]}.
\newblock \bibinfo{title}{{Hydrodynamic Gradient Expansion in Gauge Theory
  Plasmas}}.
\newblock \emph{\bibinfo{journal}{Phys Rev Lett}}
  \bibinfo{year}{2013};\bibinfo{volume}{110}(\bibinfo{number}{21}):\bibinfo{pages}{211602}.
\newblock \DOIprefix\doi{10.1103/PhysRevLett.110.211602}.
  \href{http://arxiv.org/abs/1302.0697}{\tt arXiv:1302.0697}.
\bibitem[{Denicol and Noronha(2016)}]{Denicol:2016bjh}
\bibinfo{author}{Denicol\xfnm[ G.S.]}, \bibinfo{author}{Noronha\xfnm[ J.]}.
\newblock \bibinfo{title}{{Divergence of the Chapman-Enskog expansion in
  relativistic kinetic theory}}
  \bibinfo{year}{2016};\href{http://arxiv.org/abs/1608.07869}{\tt
  arXiv:1608.07869}.
\bibitem[{Heller et~al.(2018)Heller, Kurkela, Spaliński and
  Svensson}]{Heller:2016rtz}
\bibinfo{author}{Heller\xfnm[ M.P.]}, \bibinfo{author}{Kurkela\xfnm[ A.]},
  \bibinfo{author}{Spaliński\xfnm[ M.]}, \bibinfo{author}{Svensson\xfnm[ V.]}.
\newblock \bibinfo{title}{{Hydrodynamization in kinetic theory: Transient modes
  and the gradient expansion}}.
\newblock \emph{\bibinfo{journal}{Phys Rev}}
  \bibinfo{year}{2018};\bibinfo{volume}{D97}(\bibinfo{number}{9}):\bibinfo{pages}{091503}.
\newblock \DOIprefix\doi{10.1103/PhysRevD.97.091503}.
  \href{http://arxiv.org/abs/1609.04803}{\tt arXiv:1609.04803}.
\bibitem[{Grozdanov et~al.(2019)Grozdanov, Kovtun, Starinets and
  Tadić}]{Grozdanov:2019kge}
\bibinfo{author}{Grozdanov\xfnm[ S.]}, \bibinfo{author}{Kovtun\xfnm[ P.K.]},
  \bibinfo{author}{Starinets\xfnm[ A.O.]}, \bibinfo{author}{Tadić\xfnm[ P.]}.
\newblock \bibinfo{title}{{Convergence of the Gradient Expansion in
  Hydrodynamics}}.
\newblock \emph{\bibinfo{journal}{Phys Rev Lett}}
  \bibinfo{year}{2019};\bibinfo{volume}{122}(\bibinfo{number}{25}):\bibinfo{pages}{251601}.
\newblock \DOIprefix\doi{10.1103/PhysRevLett.122.251601}.
  \href{http://arxiv.org/abs/1904.01018}{\tt arXiv:1904.01018}.
\bibitem[{Heller and Spalinski(2015)}]{Heller:2015dha}
\bibinfo{author}{Heller\xfnm[ M.P.]}, \bibinfo{author}{Spalinski\xfnm[ M.]}.
\newblock \bibinfo{title}{{Hydrodynamics Beyond the Gradient Expansion:
  Resurgence and Resummation}}.
\newblock \emph{\bibinfo{journal}{Phys Rev Lett}}
  \bibinfo{year}{2015};\bibinfo{volume}{115}(\bibinfo{number}{7}):\bibinfo{pages}{072501}.
\newblock \DOIprefix\doi{10.1103/PhysRevLett.115.072501}.
  \href{http://arxiv.org/abs/1503.07514}{\tt arXiv:1503.07514}.
\bibitem[{Romatschke(2018)}]{Romatschke:2017vte}
\bibinfo{author}{Romatschke\xfnm[ P.]}.
\newblock \bibinfo{title}{{Relativistic Fluid Dynamics Far From Local
  Equilibrium}}.
\newblock \emph{\bibinfo{journal}{Phys Rev Lett}}
  \bibinfo{year}{2018};\bibinfo{volume}{120}(\bibinfo{number}{1}):\bibinfo{pages}{012301}.
\newblock \DOIprefix\doi{10.1103/PhysRevLett.120.012301}.
  \href{http://arxiv.org/abs/1704.08699}{\tt arXiv:1704.08699}.
\bibitem[{Strickland et~al.(2018)Strickland, Noronha and
  Denicol}]{Strickland:2017kux}
\bibinfo{author}{Strickland\xfnm[ M.]}, \bibinfo{author}{Noronha\xfnm[ J.]},
  \bibinfo{author}{Denicol\xfnm[ G.]}.
\newblock \bibinfo{title}{{Anisotropic nonequilibrium hydrodynamic attractor}}.
\newblock \emph{\bibinfo{journal}{Phys Rev}}
  \bibinfo{year}{2018};\bibinfo{volume}{D97}(\bibinfo{number}{3}):\bibinfo{pages}{036020}.
\newblock \DOIprefix\doi{10.1103/PhysRevD.97.036020}.
  \href{http://arxiv.org/abs/1709.06644}{\tt arXiv:1709.06644}.
\bibitem[{Strickland(2018)}]{Strickland:2018ayk}
\bibinfo{author}{Strickland\xfnm[ M.]}.
\newblock \bibinfo{title}{{The non-equilibrium attractor for kinetic theory in
  relaxation time approximation}}.
\newblock \emph{\bibinfo{journal}{JHEP}}
  \bibinfo{year}{2018};\bibinfo{volume}{12}:\bibinfo{pages}{128}.
\newblock \DOIprefix\doi{10.1007/JHEP12(2018)128}.
  \href{http://arxiv.org/abs/1809.01200}{\tt arXiv:1809.01200}.
\bibitem[{Jaiswal et~al.(2019)Jaiswal, Chattopadhyay, Jaiswal, Pal and
  Heinz}]{Jaiswal:2019cju}
\bibinfo{author}{Jaiswal\xfnm[ S.]}, \bibinfo{author}{Chattopadhyay\xfnm[ C.]},
  \bibinfo{author}{Jaiswal\xfnm[ A.]}, \bibinfo{author}{Pal\xfnm[ S.]},
  \bibinfo{author}{Heinz\xfnm[ U.]}.
\newblock \bibinfo{title}{{Exact solutions and attractors of higher-order
  viscous fluid dynamics for Bjorken flow}}.
\newblock \emph{\bibinfo{journal}{Phys Rev}}
  \bibinfo{year}{2019};\bibinfo{volume}{C100}(\bibinfo{number}{3}):\bibinfo{pages}{034901}.
\newblock \DOIprefix\doi{10.1103/PhysRevC.100.034901}.
  \href{http://arxiv.org/abs/1907.07965}{\tt arXiv:1907.07965}.
\bibitem[{Giacalone et~al.(2019)Giacalone, Mazeliauskas and
  Schlichting}]{Giacalone:2019ldn}
\bibinfo{author}{Giacalone\xfnm[ G.]}, \bibinfo{author}{Mazeliauskas\xfnm[
  A.]}, \bibinfo{author}{Schlichting\xfnm[ S.]}.
\newblock \bibinfo{title}{{Hydrodynamic attractors, initial state energy and
  particle production in relativistic nuclear collisions}}.
\newblock \emph{\bibinfo{journal}{Phys Rev Lett}}
  \bibinfo{year}{2019};\bibinfo{volume}{123}(\bibinfo{number}{26}):\bibinfo{pages}{262301}.
\newblock \DOIprefix\doi{10.1103/PhysRevLett.123.262301}.
  \href{http://arxiv.org/abs/1908.02866}{\tt arXiv:1908.02866}.
\bibitem[{Bemfica et~al.(2018)Bemfica, Disconzi and Noronha}]{Bemfica:2017wps}
\bibinfo{author}{Bemfica\xfnm[ F.S.]}, \bibinfo{author}{Disconzi\xfnm[ M.M.]},
  \bibinfo{author}{Noronha\xfnm[ J.]}.
\newblock \bibinfo{title}{{Causality and existence of solutions of relativistic
  viscous fluid dynamics with gravity}}.
\newblock \emph{\bibinfo{journal}{Phys Rev}}
  \bibinfo{year}{2018};\bibinfo{volume}{D98}(\bibinfo{number}{10}):\bibinfo{pages}{104064}.
\newblock \DOIprefix\doi{10.1103/PhysRevD.98.104064}.
  \href{http://arxiv.org/abs/1708.06255}{\tt arXiv:1708.06255}.
\bibitem[{Bemfica et~al.(2019{\natexlab{a}})Bemfica, Disconzi and
  Noronha}]{Bemfica:2019knx}
\bibinfo{author}{Bemfica\xfnm[ F.S.]}, \bibinfo{author}{Disconzi\xfnm[ M.M.]},
  \bibinfo{author}{Noronha\xfnm[ J.]}.
\newblock \bibinfo{title}{{Nonlinear Causality of General First-Order
  Relativistic Viscous Hydrodynamics}}.
\newblock \emph{\bibinfo{journal}{Phys Rev}}
  \bibinfo{year}{2019}{\natexlab{a}};\bibinfo{volume}{D100}(\bibinfo{number}{10}):\bibinfo{pages}{104020}.
\newblock \DOIprefix\doi{10.1103/PhysRevD.100.104020}.
  \href{http://arxiv.org/abs/1907.12695}{\tt arXiv:1907.12695}.
\bibitem[{Kovtun(2019)}]{Kovtun:2019hdm}
\bibinfo{author}{Kovtun\xfnm[ P.]}.
\newblock \bibinfo{title}{{First-order relativistic hydrodynamics is stable}}.
\newblock \emph{\bibinfo{journal}{JHEP}}
  \bibinfo{year}{2019};\bibinfo{volume}{10}:\bibinfo{pages}{034}.
\newblock \DOIprefix\doi{10.1007/JHEP10(2019)034}.
  \href{http://arxiv.org/abs/1907.08191}{\tt arXiv:1907.08191}.
\bibitem[{Bjorken(1983)}]{Bjorken:1982qr}
\bibinfo{author}{Bjorken\xfnm[ J.D.]}.
\newblock \bibinfo{title}{{Highly Relativistic Nucleus-Nucleus Collisions: The
  Central Rapidity Region}}.
\newblock \emph{\bibinfo{journal}{Phys Rev}}
  \bibinfo{year}{1983};\bibinfo{volume}{D27}:\bibinfo{pages}{140--151}.
\newblock \DOIprefix\doi{10.1103/PhysRevD.27.140}.
\bibitem[{Denicol and Noronha(2018)}]{Denicol:2017lxn}
\bibinfo{author}{Denicol\xfnm[ G.S.]}, \bibinfo{author}{Noronha\xfnm[ J.]}.
\newblock \bibinfo{title}{{Analytical attractor and the divergence of the
  slow-roll expansion in relativistic hydrodynamics}}.
\newblock \emph{\bibinfo{journal}{Phys Rev}}
  \bibinfo{year}{2018};\bibinfo{volume}{D97}(\bibinfo{number}{5}):\bibinfo{pages}{056021}.
\newblock \DOIprefix\doi{10.1103/PhysRevD.97.056021}.
  \href{http://arxiv.org/abs/1711.01657}{\tt arXiv:1711.01657}.
\bibitem[{Denicol et~al.(2012)Denicol, Niemi, Molnar and
  Rischke}]{Denicol:2012cn}
\bibinfo{author}{Denicol\xfnm[ G.S.]}, \bibinfo{author}{Niemi\xfnm[ H.]},
  \bibinfo{author}{Molnar\xfnm[ E.]}, \bibinfo{author}{Rischke\xfnm[ D.H.]}.
\newblock \bibinfo{title}{{Derivation of transient relativistic fluid dynamics
  from the Boltzmann equation}}.
\newblock \emph{\bibinfo{journal}{Phys Rev}}
  \bibinfo{year}{2012};\bibinfo{volume}{D85}:\bibinfo{pages}{114047}.
\newblock \DOIprefix\doi{10.1103/PhysRevD.85.114047,
  10.1103/PhysRevD.91.039902}. \href{http://arxiv.org/abs/1202.4551}{\tt
  arXiv:1202.4551}; \bibinfo{note}{[Erratum: Phys. Rev.D91,no.3,039902(2015)]}.
\bibitem[{Pu et~al.(2010)Pu, Koide and Rischke}]{Pu:2009fj}
\bibinfo{author}{Pu\xfnm[ S.]}, \bibinfo{author}{Koide\xfnm[ T.]},
  \bibinfo{author}{Rischke\xfnm[ D.H.]}.
\newblock \bibinfo{title}{{Does stability of relativistic dissipative fluid
  dynamics imply causality?}}
\newblock \emph{\bibinfo{journal}{Phys Rev}}
  \bibinfo{year}{2010};\bibinfo{volume}{D81}:\bibinfo{pages}{114039}.
\newblock \DOIprefix\doi{10.1103/PhysRevD.81.114039}.
  \href{http://arxiv.org/abs/0907.3906}{\tt arXiv:0907.3906}.
\bibitem[{Bemfica et~al.(2019{\natexlab{b}})Bemfica, Disconzi and
  Noronha}]{Bemfica:2019cop}
\bibinfo{author}{Bemfica\xfnm[ F.S.]}, \bibinfo{author}{Disconzi\xfnm[ M.M.]},
  \bibinfo{author}{Noronha\xfnm[ J.]}.
\newblock \bibinfo{title}{{Causality of the Einstein-Israel-Stewart Theory with
  Bulk Viscosity}}.
\newblock \emph{\bibinfo{journal}{Phys Rev Lett}}
  \bibinfo{year}{2019}{\natexlab{b}};\bibinfo{volume}{122}(\bibinfo{number}{22}):\bibinfo{pages}{221602}.
\newblock \DOIprefix\doi{10.1103/PhysRevLett.122.221602}.
  \href{http://arxiv.org/abs/1901.06701}{\tt arXiv:1901.06701}.
\bibitem[{Floerchinger and Grossi(2018)}]{Floerchinger:2017cii}
\bibinfo{author}{Floerchinger\xfnm[ S.]}, \bibinfo{author}{Grossi\xfnm[ E.]}.
\newblock \bibinfo{title}{{Causality of fluid dynamics for high-energy nuclear
  collisions}}.
\newblock \emph{\bibinfo{journal}{JHEP}}
  \bibinfo{year}{2018};\bibinfo{volume}{08}:\bibinfo{pages}{186}.
\newblock \DOIprefix\doi{10.1007/JHEP08(2018)186}.
  \href{http://arxiv.org/abs/1711.06687}{\tt arXiv:1711.06687}.

\end{thebibliography}





\end{document}